\documentclass[journal]{IAENGtran}
\usepackage[ruled]{algorithm2e}
\let\labelindent\relax
\usepackage{enumitem}
\setlist[itemize]{leftmargin=*}
\usepackage{epstopdf}
\usepackage{enumerate}
\usepackage{graphicx}
\usepackage{algorithmicx}
\usepackage{multirow}
\usepackage{subfig}
\usepackage{subfloat}
\usepackage{rotating}
\usepackage{url}
\usepackage{caption}
\usepackage{array}
\usepackage{authblk}

\SetAlFnt{\small}
\SetAlCapFnt{\small}
\SetAlCapNameFnt{\small}
\SetAlCapHSkip{0pt}
\IncMargin{-\parindent}

\usepackage{etoolbox}
\apptocmd{\thebibliography}{\footnotesize}{}{}
\usepackage{url}
\usepackage[skip=1pt]{caption}
\usepackage{helvet}
\usepackage[T1]{fontenc}
\usepackage[utf8]{inputenc}
\usepackage{lipsum} 
\usepackage{fancyhdr}
\begin{document}

\title{College Student Retention: \\ When Do We Losing Them?}

\author{Mehrdad J. Bani, Mina Haji
\thanks{Manuscript received June 26, 2017; revised June 26, 2017.}
\thanks{M. J. Bani is with the Department of Computer Science, Shahid Beheshti University, Tehran, IRAN IRAN e-mail: (mehrdad4000@gmail.com).}
}

\maketitle

\begin{abstract}
One of the long term goals of any college or university is increasing the student retention.  The negative impact of student dropout are clear to students, parents, universities and society. The positive effect of decreasing studentattrition is also self-evident including higher chance of having a better career and higher
standard of life for college graduate. In view of these reasons, directors in higher education feel increasingly pressurized to outline and implement strategies to increase student retention. In this paper, we provide a detailed analysis of the student attrition problem and use statistical methods to predict when students are going to dropout from school using real case data. Our work has a number of advantages with the potential of being employed by higher education administrator of universities. We take advantage
of multiple kinds of information about different aspects of student's characteristic and
efficiently utilize them to make a personalized decision about the risk of dropout for a particular student.

\end{abstract}

\begin{IAENGkeywords}
 student retention, event prediction, statistical analysis, education decision making.
\end{IAENGkeywords}

\thispagestyle{fancy}

\vspace{-0.5em}
\section{Introduction}
One of the long-term goals of any college or university is to reduce the student attrition rate \cite{tinto1975dropout}. 
The positive effect of decreasing student attrition is also self-evident including higher chance of having a better career and higher standard of life for college graduate \cite{thomas2002student}. Not only from student perspective but also college rankings, federal funding agencies and state appropriation committees are all intrigued by student retention rates. Thus, the higher the student retention rate, the more likely that the university is positioned higher in the ranking, secure more government funds, and have easier path to program accreditations. In view of these reasons, directors in higher education feel increasingly pressurized to outline and implement strategies to increase student retention.

The study of college student retention is very important for educational researchers, managers, and the higher education members.
Costs accrue to society, the institution, and the student when degree completion is not realized \cite{pascarella2005college}.
From a societal perspective, the achievement of a college diploma improve mediate background resources, like family economic situation. Thus, it impacts on subsequent occupational status, potential earnings, and social status achievement \cite{thomas2002student}. This fact is proved by the difference between social situation achievements of every single person from the same level of economic condition with different levels of educational degree \cite{pascarella2005college}. 
From the institution’s perspective, maintaining enrollments is important to economic stability. As the number of new students has fluctuated, so finding the characteristics of students that cause to remain enroll and graduate is very important. Universities and colleges and all other institutions become fully aware of this fact that the primary assets needed to recruit, enroll, enlist, register, advise and assist a new student are the same whether that student remains and graduates or not. For instance, publication and marketing expenses, costs associated with maintaining a staff of professional counselors, travel costs connected with conducting college fairs and informational meetings, costs related with tele-counseling, staff time contacting students for yield enhancement, and staff time conducting academic advising, are all examples of pre-enrollment expenses experienced by the institution. 

In higher education, student retention rate can be defined as the percentage of students who after completing a semester, return to the same university for the following semester. Student retention rate not only has effect on the university but also affect the nearby cites. An institution’s retention rate influences job opportunities for students and public opinions. Universities are eager to find out which factors or attributes are important for their students’ retention, how they can address this issue and with help of this kind of analysis they can improve their retention rate. It is important because higher retention rate will increase university funding from state and also help them to recruit more students and faculty in order to give better service to community. College student graduation rates are often used as a measure of institution's performance. However, on the other hand the dropout rate can show university failure to persuade student finish their school. These kind of analysis play a significant role for universities decision about how to expand their majors' capacities and how to allocate financial aid between students. It is not beneficial to give financial aid to who are at high risk of drop out. More important that universities are accountable for attrition rate on their colleges so they will be penalized with narrowing their fund. In recent years the cost of higher education is increased continuously and amount of fund from state and federal government that colleges received is decreased so universities should try to spend their fund very wisely and doing this without analytical studies is not feasible \cite{fard2016early}. 
From university point of view maintaining enrollment is important for economic stability, institutional success and managing resource allocation which university growth. From student perspective, it will result in higher chance of graduation that cause higher chance of getting good job, earn more money and better life style. In general, knowing the reasons for student dropout can help the faculties and administrators to take necessary actions so that the success percentage can be improved.

Event prediction is an important area of research where the goal is to predict the occurrence of an event \cite{mahtabIEOM2015}.
In higher education, many modeling techniques were found to help educational institutions to predict at-risk students \cite{nandeshwar2011learning}. This results in planning for interventions and better understanding and addressing fundamental issues that cause the student attrition problem. 
In the past decades, comprehensive models have been developed to address the college student attrition problem. Most of the earlier studies try to understand the reasons behind student dropout by developing theoretical models \cite{tinto1987leaving}.  
For many years, statistical methods have been used widely to predict student dropout and also find the important factors that has effect on that \cite{zhang2004identifying, jones2010redefining}. Regression is one of the primary techniques that has been applied in this area \cite{dey1993statistical}. 
Traditional methods such as regression have been used to identify dropout students for decades  \cite{deberard2004predictors, lin2009student}. Recently, due to the successful result of implementing machine learning methods in other area such as healthcare \cite{jahanbani2016computational} and robotics \cite{fard2016machine}, researchers in the area of machine learning and data mining, tried to address the student retention phenomenon as well \cite{delen2011predicting,yadav2012mining,thammasiri2014critical}. Genetic algorithms for selecting feature subset and artificial neural networks for performance modeling have been developed to give better prediction of first year high risk students to dropout at Virginia Commonwealth University \cite{alkhasawneh2011developing}. Several classification algorithms including Bayes classifier \cite{pandey2011data,bhardwaj2012data},decision tree \cite{herzog2006estimating,quadri2010drop,yu2010data}, boosting methods and support vector machines \cite{zhang2010using, lakkaraju2015machine} have been developed to predict student attrition with higher accuracy compared to the traditional statistical methods. In spite of the success of survival analysis methods in other domains such as healthcare \cite{7564399}, engineering \cite{doi:10.1108/IJQRM-05-2011-0075}, etc., there is only a limited attempt of using these methods in student retention problem \cite{ameri2016survival, ishitani2003longitudinal}.

While most of the previous works focused on predicting dropout student, in this paper we try to answer the important question of "when a student is going to dropout?" In another word,  we implement different statistical methods on real student retention data to predict when a student is going to dropout from school based on the pre-school data which is available at the beginning of his/her study. The fundamental idea is that we can utilize the survival analysis method at an early stage of college study to predict student dropouts.

\section{Statistical Method}
The primary goal of this work is to implement different statistical and machine learning method to predict when a student is going to dropout based on both pre-school attributes.   

\subsection{Linear Regression}
Regression models are among the most important methods in predictive analytics \cite{draper1966applied}. Linear regression was the first type of regression analysis to be studied rigorously, and to be used extensively in practical applications. Regression models are primary established to model a mathematical equation to represent the interactions between the different variables in consideration. Depending on the situation, there are a wide variety of models that can be applied. The mainstay one is linear regression which tries to find the linear relation between dependent variable and a set of independent or predictor variables.
\begin{equation}
y=w_0+\sum_{k=1}^p w_kx_k
\end{equation}
where $w$s are coefficients. The goal here is to select the parameters of the model so as to minimize the sum of the squared error. In student retention problem, it was one of the primary techniques that has been tested \cite{dey1993statistical}. It can be applied to model time-to-event problem which in student dropout data is time to dropout. 

\subsection{Support Vector Regression}
Support Vector Regression (SVR) is a sub-category of Support Vector Machine (SVM) where it can solve regression problems \cite{drucker1997support}. The model produced by Support Vector Regression depends only on a subset of the training data. Thus, training the original SVR means solving
\[ \mbox{minimize\ } \frac{1}{2}|w|^{2} \] 
having these two constraints,
$$\begin{cases}
y_i - \langle w, x_i \rangle - b \le \epsilon \\
\langle w, x_i \rangle + b - y_i \le \epsilon
\end{cases}$$
where in the above optimization problem $x_i$ is a training sample with target value $y_i$ and $\epsilon$ is a free parameter for a threshold.

\subsection{Cox Regression}
One of the popular methods in survival analysis is the Cox proportional hazard model \cite{cox1972regression}. The Cox regression model is a semi-parametric technique which has fewer assumptions than typical parametric methods. In particular, and in contrast with parametric models, it makes no assumptions about the shape of the baseline hazard function. The Cox model provides a useful and easy way to interpret information regarding the relationship of the hazard function to predictors. The hazard function for the Cox proportional hazard model has the form
\begin{equation}
h(t|X) = h_0(t)\exp(\beta_1X_1 + \cdots + \beta_pX_p) = h_0(t)e^{(\beta X)}
\label{cox}
\end{equation}
where $h_0(t)= e^{\alpha(t)}$ is the baseline hazard function at time $t$ and $\exp(\beta_1X_1 + \cdots + \beta_pX_p)$ is the risk associated with the covariate values.
Therefore, the survival probability function for Cox model can be formulated as
\begin{equation}
S (t \mid X)=S_0(t) ^{{exp}({\beta X}) }
\label{survcox}
\end{equation}
where
\begin{equation}
S_0 (t)=e^{-\int_0^t h_0(x) dx}
\label{S0cox}
\end{equation}

Parameters of the Cox regression model are estimated by maximizing the partial likelihood.
Based on Cox regression formula, a partial likelihood can be constructed from the dataset as follows:
\begin{equation}
L(\beta) = \prod_{i:\delta_i=1}\frac{\theta_i}{\sum_{j:t_j\ge t_i}\theta_j}
\label{likeli}
\end{equation}
where $\theta_i = \exp(\beta X_i)$ and $(X_1, ..., X_n)$ are the covariate vectors for the $n$ independently sampled individuals in the dataset. By solving $\frac{\partial L(\beta)}{\partial \beta}$=0, the covariate coefficient can be estimated as $\hat{\beta}$. To obtain the baseline hazard function, in full likelihood function, $\beta$ should be replaced by $\hat{\beta}$. Thus, $h_0(t_i)$ can be obtained
\begin{equation}
\hat{h}_0\big(t_{(i)}\big)=\frac{1}{\sum_{j\in R({t_{(i)}})}\theta_j}
\label{e5}
\end{equation}

\section{Results and Discussion}
In this study, a dataset was compiled by tracking 5000 students enrolled at a college in our home town starting from 2008 until 2015 from different department.
After the data preparation and the necessary pre-processing, we ended up with 16 predictor attributes; Gender, Ethnicity, Marital status, Residence county, Student's income, Father's income, Mother's income, Number of household, High School GPA, Reading Score, Math Score, Science Score, High school graduation age, Age of admission, Student's college, Student's major

In order to have a quantitative measure for estimating time to dropout which make the comparison between different methods more straightforward.
We divide our data into training and testing sets. We report the results of both 5-fold cross validation on training set and the test data in separate tables. In the first one, we use standard technique of stratified 5-fold cross validation. The evaluation metrics for each method is then computed as an average of the ten experiments. To assess the performance of the different methods, following metrics are used:

\begin{itemize}
\item{\textit{{Mean Absolute Error (MAE)}}} is a quantity used to measure how close the predictions are to the actual outcomes. The mean absolute error is given by
$$MAE = \frac{1}{n}\sum_{i=1}^n \left|\hat{y_i} - y_i\right|$$
where $\hat{y_i}$ is the predicted value and $y_i$ is the true value for subject $i$.
\end{itemize}
MAE treats both underestimating and overestimating of actual value in the same manner. However, in student retention problem, these types of errors have different meaning. For example, any model that has the ability to predict the semester of dropout earlier than the actual semester has more value because, in this case, an individualized intervention programs might help to reduce the student dropout rate. Therefore, we also evaluated the models using the following two domain based metrics:


\begin{itemize}
\item{\textit{{Underestimated Prediction Error Rate (UPER)}}} is defined as the fraction of the underestimated prediction output over the entire prediction error \cite{ameri2016survival}.

$$UPER=\frac{\sum_{i=1}^n I(\hat y_i < y_i)}{\sum_{i=1}^n I(\hat y_i < y_i)+\sum_{i=1}^n I(\hat y_i > y_i)}$$

\item{\textit{{Overestimated Prediction Error Rate (OPER)}}}: since the total number of error is a constant, OPER can be calculated as \cite{ameri2016survival}
$$OPER=1-UPER$$
\end{itemize}

It should be noted that, in the student retention problem, any model with higher UPER than OPER will be of great interest because it is better to underestimate the semester of dropout earlier, rather than overestimate it.

One of the primary objectives of this work, is to build a model to estimate the semester of dropout at the beginning of the study. As discussed earlier, one of the drawbacks of using linear regression in the presence of censored data is that this information cannot be handled properly thus resulting in a biased estimation of time to dropout for the student retention problem. Therefore, the standard classification and regression methods will not be able to answer the important question of ``when a student is going to dropout?" in the presence of censored data. In this paper, we apply our survival analysis based framework to answer this question. Table \ref{tab5} shows the result of 10-fold cross-validation training data and 2009 data as test data, using first and second experimental setups. We compare the result of {\emph Cox} with linear regression and well-known Support Vector Regression (SVR) \cite{drucker1997support}. We should also mention that {\em TD-Cox} cannot be used for this purpose as we only want to use pre-enrollment information to estimate the semester of dropout. {\em TD-Cox} uses semester-wise information which are available only after the students begin their semester. In other words, we are interested in estimating the semester of dropout without using any semester-wise information (After-enrollment variables).

From our results, we can conclude that the {\emph{Cox}} model outperforms other methods. From Table \ref{tab5}, it is clear that, in the presence of censored data, survival based methods such as {\em Cox} model have a better performance compared to the traditional regression methods. We can also observe that {\em Cox} has the higher UPER, which indicates that majority of errors come from underestimating the semester of dropout. This will allow us to have a better individualized intervention programs with more focus towards specific high-risk students as early as (s)he starts the school.
Consequently, we are able to maximize the retention rate which can then translate into increasing number of graduations from the university. 

\vspace{1em}

\begin{table}[htbp]
\centering
\caption{\small{Comparison between different method for predicting "when student is going to dropout?".} 
\label{tab5}}
\begin{tabular}{|c|c|c|c|c|c|c|}
\cline{2-7} 
\multicolumn{1}{c|}{} & \multicolumn{3}{c|}{\textbf{5-fold CV}} & \multicolumn{3}{c|}{\textbf{Test}}\tabularnewline
\hline 
\textbf{Model} & \textbf{MAE } & \textbf{UPER} & \textbf{OPER} & \textbf{MAE } & \textbf{UPER} & \textbf{OPER}\tabularnewline
\hline 
\textbf{Regression} & 1.45 & 0.698 & 0.302 & 1.53 & 0.659 & 0.341\tabularnewline
\hline 
\textbf{SVR} & 1.32 & 0.711 & 0.289 & 1.39 & 0.694 & 0.306\tabularnewline
\hline 
\textbf{Cox} & \textbf{1.28} & \textbf{0.725} & \textbf{0.275} & \textbf{1.35} & \textbf{0.711} & \textbf{0.289}\tabularnewline
\hline 
\end{tabular}
\end{table}%

This study has shown the benefits of survival analysis as a methodology for the study of college student dropout behaviors. 
This can help colleges and universities to design effective retention strategies to help students to complete their degree before dropping out from school. 
Hence, our work will enable universities to utilize their resources more efficiently by targeting only the high risk students who are more vulnerable of dropping out of their study at any given semester.

\vspace{0.2in} 

\vspace{-2.4em}
\section{Conclusion}
Predicting students who will dropout from their study is an important and challenging task for academic institutions. However, little research in higher education has focused on the employment of data mining and statistical methods for predicting retention. 
Benefits of survival analysis as an approach for studying the timing of events are clear in many different application domains which deal with longitudinal data. In this paper, we develop a survival analysis based framework on the problem of estimating when students are dropping out from their study during their early stage of higher educational life.
Our study has shown the benefits of survival analysis as a methodology for the study of college student dropout behaviors. It should be noted that the majority of dropouts happen during freshman year (first two semesters). Thus, the ability to build a model that provides the prediction result at an early stage with high accuracy is very crucial.
Our results motivate that the statistical method allows educational institutions to undertake timely measures and actions in their student attrition problem. Based on the findings of this paper, we can use pre-enrollment information as screening test to identify students with higher risk of dropout. It is recommended that future research on student retention behaviors should be conducted using other available information such as course interaction website which contains student activity information for each course. This can help with developing interventions that can be deployed early on in a course to improve student success within the course, and in turn, retention.

\bibliographystyle{BibTeXtran}
\bibliography{myref}

\end{document}